# Wireless Information and Power Transfer in Full-Duplex Communication Systems

Alexander A. Okandeji, Muhammad R. A. Khandaker, Kai-Kit Wong
Dept. of Electronic and Electrical Engineering
University College London
Gower Street, London, WC1E 7JE, United Kingdom
Email: alexander.okandeji.13@ucl.ac.uk, m.khandaker@ucl.ac.uk, kai-kit.wong@ucl.ac.uk

*Abstract*—This paper considers the problem of maximizing the sum-rate for simultaneous wireless information and power transfer (SWIPT) in a full-duplex bi-directional communication system subject to energy harvesting and transmit power constraints at both nodes. We investigate the optimum design of the receive power splitters and transmit powers for SWIPT in full-duplex mode. Exploiting rate-split method, an iterative algorithm is derived to solve the non-convex problem. The effectiveness of the proposed algorithm is justified through numerical simulations.

## I. Introduction

The possibility of the simultaneous wireless transfer of information and power within the same network has been recognised as a promising energy harvesting technique. The idea of transmitting information and energy simultaneously was first proposed by varshney in [1]. In order to describe the fundamental trade-off between the rates at which energy and dependable information is transmitted over a noisy channel, the authors in [1] proposed a capacity-energy function. Wireless information and power transfer subject to co-channel interference was studied in [2], in which optimal designs to achieve different outage-energy trade-offs as well as rate-energy trade-offs were derived. Energy harvesting receiver in multiple-input multiple-output (MIMO) relay systems were studied in [3]. The authors in [3] determined different trade-offs between the energy transfer and information rate for a joint optimal source and relay precoders. SWIPT for multi-user systems was studied in [4]. It was shown in [4] that for multiple access channels with received energy constraints, time sharing is necessary to achieve the maximum sum-rate when received energy constraints is sufficiently large.

Recently, there has been an upsurge of interest in full-duplex (FD) communication due to the fact that full-duplexity can offer higher spectral efficiency compared to its half-duplex counterpart. Transmit strategies for a full-duplex point-to-point system with residual self-interference were studied in [5]. The authors in [5] analysed FD system constraints at optimality and thus developed power adjustment schemes which maximizes the system sum-rate in different scenarios.

However, the residual self-interference (RSI) resulting from the concurrent transmission and reception at the same node raises the noise floor and is a dominant factor in the performance of full-duplex communication systems. Hence considerable efforts have been made in mitigating the effects of RSI in FD systems. Digital self-interference cancellation technique for FD wireless system was studied in [6]. It was shown in [6] that the average amount of self-interference cancellation achieved for antenna separation and digital cancellation at 20 cm and 40 cm spacing between interfering antennas was 70 dB and 76 dB, respectively. The RSI has been modeled by additive white Gaussian noise (AWGN) with zero-mean and known variance in some works as well [7].

A characterization of the effect of increasing self-interference power on the amount of active cancellation and rate has not been reported in the aforementioned works. The authors in [8] presented an experiment-based characterization of passive suppression and active self-interference cancellation mechanisms in full-duplex wireless communication systems. It was shown in [8] that the average amount of cancellation increases for active cancellation techniques as the received self-interference power increases. Based on extensive experiments, the authors in [8] showed that a total average cancellation of 74 dB can be achieved.

However, for all the reported full-duplex systems, it has been observed that self-interference cancellation cannot suppress the self-interference down to the noise floor [8], [9]. A sophisticated digital self-interference cancellation technique has been proposed in [9] that eliminates all transmitter impairments, and significantly mitigates the receiver phase noise and nonlinearity effects. The proposed technique in [9] significantly mitigates the self-interference signal to $\sim 3$ dB higher than the receiver noise floor, which results in up to $67 - 76\%$ rate improvement compared to conventional half-duplex systems at 20 dBm transmit power values.

More recently, interest have been focussed on the study of SWIPT in full-duplex systems as it has the potential to improve spectral efficiency and achieve simultaneous transmission of information and power [10], [11]. The authors of [10] considered an access point operating in FD mode that broadcasts wireless energy to a set of distributed users in the downlink and, at the same time, receives independent information from the users via time-division multiple access in the uplink. In contrast, a scenario is considered in [11] where an energy-constrained FD relay node assists the information transmission from the source to the destination using the energy harvested from the source.

In this paper, we consider SWIPT in a point-to-point full-duplex wireless communication system for simultaneous bidirectional communication where two nodes equipped with two antennas, one used for signal transmission and the other used for signal reception, communicate in full-duplex mode. Our aim is to maximise the end-to-end sum-rate for SWIPT in FD system while maintaining the energy harvesting threshold at each node by optimizing the receive power splitter and transmit power at each node. Due to insufficient knowledge of the self-interfering channel, we consider the worst-case based model where the magnitude of the estimation error is bounded. Since the problem is strictly non-convex, we propose an alternating solution. In particular, we show that for fixed power splitting ratio, the optimal transmit powers can be obtained by introducing a rate-split scheme between the two nodes, whereas for given transmit powers, closed-form expressions for power splitting ratios can be derived. Numerical simulations are carried out to demonstrate the performance of the proposed scheme.

The rest of this paper is organized as follows. In Section II, the system model of a full-duplex point-to-point communication network with power splitting based energy harvesting nodes is introduced. The proposed joint transmit power and receive PS ratio design algorithm is developed in Section III. Section IV shows the simulation results under various scenarios. Conclusions are drawn in Section V.

## II. SYSTEM MODEL AND PROBLEM FORMULATION

Let us consider a full-duplex point-to-point wireless communication system as illustrated in Fig. 1. It is assumed that each node houses identical transmitter-receiver pair. Each receiver intends to simultaneously decode information and harvest energy from the received signal. Let us define the received signal at node 1 and node 2 as $y_{2 \to 1}$ and $y_{1 \to 2}$, respectively. Let us also denote the transmit and receive antennas at nodes 1 and 2 by $(a,b)$ and $(c,d)$, respectively. Thus, the received signal at node 1 is given by

$$y_{2 \to 1} = h_{cb}x_2 + h_{ab}x_1 + n_{A_1} \tag{1}$$

and the received signal at node 2 is

$$y_{1 \to 2} = h_{ad}x_1 + h_{cd}x_2 + n_{A_2} \tag{2}$$

where $x_1$ denotes the transmitted signal from node 1 to node 2, $x_2$ denotes the transmitted signal from node 2 to node 1, $h_{ad}$ denotes the wireless channel from node 1 to node 2, $h_{cb}$ denotes the wireless channel from node 2 to node 1, $h_{ab}$ and $h_{cd}$ denote the self-interference channel at node 1 and node 2, respectively. $n_{A1}$ and $n_{A2}$ are defined as the AWGN with zero mean and unit variance at the receiver antenna at node 1 and node 2, respectively.

In this paper, we assume that each receiver is equipped with a power splitting device which coordinates the processes of information decoding and energy harvesting from the received signal. For simplicity, we assume that the received signal is split such that a $\rho_k, k = 1, 2$, portion of the signal power is fed to the information decoder (ID) and the remaining $1 - \rho_k, k = 1, 2$, is fed to the energy harvester (EH) at node $k$. Thus the

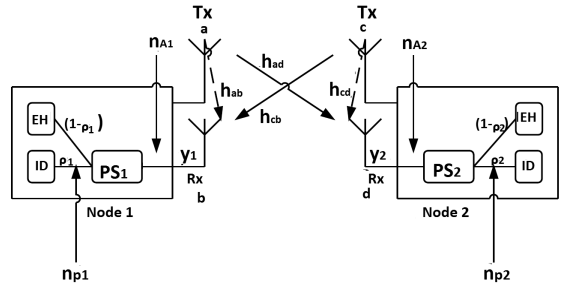

Fig. 1. Energy harvesting full-duplex communication system.

signal split to the ID of node 1 and 2 are given, respectively, by

$$y_{2 \to 1}^{ID} = \sqrt{\rho_1}(h_{cb}x_2 + h_{ab}x_1 + n_{A_1}) + n_{p1} \tag{3a}$$

$$y_{1 \to 2}^{ID} = \sqrt{\rho_2}(h_{ad}x_1 + h_{cd}x_2 + n_{A_2}) + n_{p2} \tag{3b}$$

where $n_{pk}$, $k = 1, 2$, is the noise introduced by the RF band to baseband signal conversion operation and is defined as $n_{pk} \sim CN(0, \sigma_p^2)$, $k = 1, 2$. Also, the signal split to the EH is expressed as

$$y_{1 \to 2}^{EH} = \sqrt{1 - \rho_2}(h_{ad}x_1 + h_{cd}x_2 + n_{A_2}) \tag{4a}$$

$$y_{2 \to 1}^{EH} = \sqrt{1 - \rho_1}(h_{cb}x_2 + h_{ab}x_1 + n_{A_1}). \tag{4b}$$

Several power splitting schemes have been considered in the literature including: i) Uniform power splitting (UPS), where equal power is split between the ID and EH ($\rho_k = \frac{1}{2}$); ii) On-Off power splitting (OOPS), where $\rho_k \in (0, 1)$; iii) optimal power splitting (OPS) schemes [12]. In OOPS, depending on the system condition, the receiving nodes can switch between ID and EH modes which is sometimes referred to as *time-switching*. In time switching, when the information receiver is switched to its off mode, all signal power is used for energy harvesting and vice versa.

Thus the power harvested at the EH of node 1 and node 2 is thus given by

$$Q_{1 \to 2} = \alpha(1 - \rho_2)(E[|h_{ad}x_1 + h_{cd}x_2 + n_{A_2}|^2]) \tag{5a}$$

$$Q_{2 \to 1} = \alpha(1 - \rho_1)(E[|h_{cb}x_2 + h_{ab}x_1 + n_{A_1}|^2]) \tag{5b}$$

where $\alpha$ denotes the energy conversion efficiency of the EH at each receiver that accounts for the loss in the energy transducer for converting the harvested energy to electrical energy to be stored. In practice, an energy harvesting circuit is equipped at the energy harvesting receiver which is used to convert the received RF power into direct current (DC) power. It has been proved that the efficiency of diode-based energy harvesters is non-linear and largely depends on the input power level [13]. Hence, the energy conversion efficiency ($\alpha$) should be included in optimization expressions. However, for simplicity, it is assumed that $\alpha = 1$ in this paper.

Let us now define $C_{1 \to 2}$ and $C_{2 \to 1}$ as the information rate from node $1 \to 2$ and node $2 \to 1$, respectively. Thus

the instantaneous capacity at node 1 and 2 can be written, respectively, as

$$C_{2\to 1} = \log_2\left(1 + \frac{\rho_1|h_{cb}|^2 P_2}{\rho_1(|h_{ab}|^2 P_1 + \sigma_{A_1}^2) + \sigma_{p_1}^2}\right) \quad (6a)$$

$$C_{1\to 2} = \log_2\left(1 + \frac{\rho_2|h_{ad}|^2 P_1}{\rho_2(|h_{cd}|^2 P_2 + \sigma_{A_2}^2) + \sigma_{p_2}^2}\right). \quad (6b)$$

We show later in Section IV through numerical results that if the residual self-interference is not handled properly, it dominates the system performance and prevents from exploiting the benefits of full-duplex by decreasing the information rate. Considering the fact that the RSI can not be eliminated completely, we consider the worst-case performance based on deterministic model for imperfect self-interfering channels. In particular, we assume that the self-interference channels $h_{ab}$ and $h_{cd}$ lie in the neighbourhood of the estimated channels $\hat{h}_{ab}$ and $\hat{h}_{cd}$, respectively, available at the nodes. Thus the actual channels due to imperfect self-interference channel estimate can be written as

$$h_{ab} = \hat{h}_{ab} + \triangle h_{ab} \quad (7a)$$
$$h_{cd} = \hat{h}_{cd} + \triangle h_{cd} \quad (7b)$$

where $\triangle h_{ab}$ and $\triangle h_{cd}$ represents the channel uncertainties which are assumed to be bounded such that

$$|\triangle h_{ab}| = |h_{ab} - \hat{h}_{ab}| \leq \epsilon_1 \quad (8a)$$
$$|\triangle h_{cd}| = |h_{cd} - \hat{h}_{cd}|_2 \leq \epsilon_2 \quad (8b)$$

for some $\epsilon_1 \geq 0$ and $\epsilon_2 \geq 0$, where $\epsilon_k$, $k = 1, 2$, depends on the accuracy of the CSI estimates. To efficiently define the worst-case self-interference level, we modify (7a) and (7b) using the triangle inequality [14]. It follows from (7a) that

$$|h_{ab}|^2 = |(\hat{h}_{ab} + \triangle h_{ab})|^2 \leq |\hat{h}_{ab}|^2 + |\triangle h_{ab}|^2 \leq |\hat{h}_{ab}|^2 + \epsilon_1^2. \quad (9)$$

Note that $\epsilon_1$ is the minimal knowledge of the upper-bound of the channel error which is sufficient enough to describe the error in the absence of statistical information about the error. Thus from (9), we obtain

$$\max_{|\triangle h_{ab}| \leq \epsilon_1} |(\hat{h}_{ab} + \triangle h_{ab})|^2 \leq |\hat{h}_{ab}|^2 + \epsilon_1^2. \quad (10)$$

Similar results can be obtained from (8b) as

$$\max_{|\triangle h_{cd}| \leq \epsilon_2} |(\hat{h}_{cd} + \triangle h_{cd})|^2 \leq |\hat{h}_{cd}|^2 + \epsilon_2^2. \quad (11)$$

On the other hand, it holds that

$$|(\hat{h}_{ab} + \triangle h_{ab})|^2 \geq |\hat{h}_{ab}|^2 - |\triangle h_{ab}|^2 \geq |\hat{h}_{ab}|^2 - \epsilon_1^2 \quad (12)$$

and

$$|(\hat{h}_{cd} + \triangle h_{cd})|^2 \geq |\hat{h}_{cd}|^2 - |\triangle h_{cd}|^2 \geq |\hat{h}_{cd}|^2 - \epsilon_2^2. \quad (13)$$

Here, we assume that $|\hat{h}_{ab}| \geq |\triangle h_{ab}|$ and $|\hat{h}_{cd}| \geq |\triangle h_{cd}|$ which essentially means that the errors $|\triangle h_{ab}|$ and $|\triangle h_{cd}|$ are sufficiently small. Accordingly,

$$\min_{|\triangle h_{ab}| \leq \epsilon_1} |(\hat{h}_{ab} + \triangle h_{ab})|^2 \geq |\hat{h}_{ab}|^2 - \epsilon_1^2 \quad (14)$$

and

$$\min_{|\triangle h_{cd}| \leq \epsilon_2} |(\hat{h}_{cd} + \triangle h_{cd})|^2 \geq |\hat{h}_{cd}|^2 - \epsilon_2^2. \quad (15)$$

Substituting the results obtained in (14) and (15) into (5a) and (5b), respectively, we have the minimum power harvested at the EH of node 1 and node 2 given by

$$\min_{|\triangle h_{ab}| \leq \epsilon_1} Q_{2\to 1}$$
$$\geq (1-\rho_1)(|h_{cb}|^2 P_2 + (|\hat{h}_{ab}|^2 - \epsilon_1^2)P_1 + \sigma_{A_1}^2) \quad (16a)$$
$$\min_{|\triangle h_{cd}| \leq \epsilon_2} Q_{1\to 2}$$
$$\geq (1-\rho_2)(|h_{ad}|^2 P_1 + (|\hat{h}_{cd}|^2 - \epsilon_2^2)P_2 + \sigma_{A_2}^2). \quad (16b)$$

The minimum instantaneous capacity at node 1 and 2 can be written, respectively, as

$$\min_{|\triangle h_{ab}| \leq \epsilon_1} C_{2\to 1}$$
$$\geq \log_2\left(1 + \frac{\rho_1|h_{cb}|^2 P_2}{\rho_1((|\hat{h}_{ab}|^2 + \epsilon_1^2)P_1 + \sigma_{A_1}^2) + \sigma_{p_1}^2}\right) \quad (17a)$$
$$\min_{|\triangle h_{cd}| \leq \epsilon_2} C_{1\to 2}$$
$$\geq \log_2\left(1 + \frac{\rho_2|h_{ad}|^2 P_1}{\rho_2((|\hat{h}_{cd}|^2 + \epsilon_2^2)P_2 + \sigma_{A_2}^2) + \sigma_{p_2}^2}\right). \quad (17b)$$

The sum-rate of information across the communication system is given by

$$R_{sum} \triangleq C_{2\to 1} + C_{1\to 2}$$
$$= \log_2\left(1 + \frac{\rho_2|h_{ad}|^2 P_1}{\rho_2(|\hat{h}_{cd}|^2 + \epsilon_2^2)P_2 + \rho_2\sigma_{A_2}^2 + \sigma_{p_2}^2}\right)$$
$$+ \log_2\left(1 + \frac{\rho_1|h_{cb}|^2 P_2}{\rho_1(|\hat{h}_{ab}|^2 + \epsilon_1^2)P_1 + \rho_1\sigma_{A_1}^2 + \sigma_{p_1}^2}\right). \quad (18)$$

In order to maximise the sum-rate of SWIPT in full-duplex system, the optimal transmit power and receive power splitting problem with transmit power and harvested energy constraints at node 1 and node 2 can be formulated as

$$\max_{\rho_1, \rho_2 \in (0,1), P_1, P_2} R_{sum} \quad (19a)$$
$$\text{s.t.} \quad \min_{|\triangle h_{ab}| \leq \epsilon_1} Q_{2\to 1} \geq \bar{Q}_{2\to 1} \quad (19b)$$
$$\min_{|\triangle h_{cd}| \leq \epsilon_2} Q_{1\to 2} \geq \bar{Q}_{1\to 2} \quad (19c)$$
$$0 \leq P_1 \leq P_{max} \quad (19d)$$
$$0 \leq P_2 \leq P_{max} \quad (19e)$$

where $\bar{Q}_{1\to 2}$ and $\bar{Q}_{2\to 1}$ are the minimum amount of harvested energy required to maintain the receivers operation, and $P_{max}$ is the maximum available transmit power budget at node 1 and node 2, respectively.

## III. PROPOSED SOLUTION

In this section we address the optimum design of the receive power splitter and transmit power for SWIPT in full-duplex communication systems assuming that the instantaneous CSI is known at the transmitter.

Since the problem (19) is non-convex, it is very difficult to obtain a closed-form solution that jointly optimizes $\rho_1$, $\rho_2$, $P_1$, and $P_2$. Hence, to solve this problem, we propose a two-step iterative process. First, we fix the splitter coefficients, i.e., $\rho_1, \rho_2 \in (0,1)$ and obtain the optimal values for $P_1$ and $P_2$. Then, we use the optimal $P_1^*$ and $P_2^*$ to obtain optimal $\rho_1^*$ and $\rho_2^*$.

### A. Transmit Power Optimization

Even for fixed $\rho_1, \rho_2$, the problem is still non-convex since the objective function is not a concave function. Hence to efficiently solve problem (19), we first transform it using the idea of the rate-split method [15], formulated as

$$\max_{P_1, P_2} \quad R_{sum} \tag{20a}$$
$$\text{s.t.} \quad \min_{|\Delta h_{ab}| \leq \epsilon_1} C_{2 \to 1} \geq \eta R_{sum} \tag{20b}$$
$$\min_{|\Delta h_{cd}| \leq \epsilon_2} C_{1 \to 2} \geq (1-\eta) R_{sum} \tag{20c}$$
$$\min_{|\Delta h_{ab}| \leq \epsilon_1} Q_{2 \to 1} \geq \bar{Q}_{2 \to 1} \tag{20d}$$
$$\min_{|\Delta h_{cd}| \leq \epsilon_2} Q_{1 \to 2} \geq \bar{Q}_{1 \to 2} \tag{20e}$$
$$0 \leq P_1 \leq P_{max} \tag{20f}$$
$$0 \leq P_2 \leq P_{max}, \tag{20g}$$

where $\eta \in [0,1]$. For any given $\eta$, the first two constraints typically impose a rate-split between the two nodes i.e., $\eta$ is a rate-split scheme. If we can solve (20) to get $R_{sum}(\eta)$ for given $\eta$, then we can do a one-dimensional search on $\eta$ to find the maximum $R_{sum}(\eta^o)$ under the optimal rate-split scheme $\eta^o$. To proceed, let us first rewrite the optimization problem (20) as

$$\max_{P_1, P_2} \quad r \tag{21a}$$
$$\text{s.t.}$$
$$\rho_2 |h_{ad}|^2 P_1 \geq \beta_1 \left( \rho_2(|\hat{h}_{cd}|^2 + \epsilon_2^2) P_2 + \rho_2 \sigma_{A_2}^2 + \sigma_{p_2}^2 \right) \tag{21b}$$
$$\rho_1 |h_{cb}|^2 P_2 \geq \beta_2 \left( \rho_1(|\hat{h}_{ab}|^2 + \epsilon_1^2) P_1 + \rho_1 \sigma_{A_1}^2 + \sigma_{p_1}^2 \right) \tag{21c}$$
$$(1-\rho_1)(|h_{cb}|^2 P_2 + (|\hat{h}_{ab}|^2 - \epsilon_1^2) P_1 + \sigma_{A_1}^2) \geq \bar{Q}_{2 \to 1} \tag{21d}$$
$$(1-\rho_2)(|h_{ad}|^2 P_1 + (|\hat{h}_{cd}|^2 - \epsilon_2^2) P_2 + \sigma_{A_2}^2) \geq \bar{Q}_{1 \to 2} \tag{21e}$$
$$0 \leq P_1 \leq P_{max} \tag{21f}$$
$$0 \leq P_2 \leq P_{max}, \tag{21g}$$

where $r$ is the optimal objective value for problem (20), $\beta_1 = 2^{\eta r} - 1$, and $\beta_2 = 2^{(1-\eta)r} - 1$. Problem (21) is convex and can be efficiently solved by the disciplined convex programming toolbox like CVX [16]. After solving problem (21), the optimal values of the transmit power at node 1 and node 2 denoted as $P_1^*$ and $P_2^*$, respectively, gives the optimal achievable sum-rate $r^o$ at fixed values of $(\rho_1, \rho_2) \in (0,1)$.

Algorithm 1 summarises the whole procedure of solving problem (21). We can see that in both initialization and optimization steps, solving for $r$ is the elementary operation in each iteration. We use CVX package to solve the problem and iteratively update $r$ by using the bisection method. The bounds of the rate search interval are obtained as follows. The lower bound $r_{low}$ of the rate search is obviously 0 while the upper bound $r_{max}$ is defined as the achievable sum-rate at zero RSI.

---

**Algorithm 1** Procedure for solving problem (21)

1: Fix $\rho_1$ and $\rho_2$ such that $\rho_1, \rho_2 \in (0,1)$.
2: Set $\eta(0) = 0$.
3: At step k, set $\eta(k) = \eta(k-1) + \Delta\eta$ until $\eta(k) = 1$, where $\Delta\eta$ is the searching step size.
4: Initialize $r_{low} = 0$, and $r_{up} = r_{max}$.
5: Repeat
   1) Set $r \leftarrow \frac{1}{2}(r_{low} + r_{up})$ and calculate $\beta_1$, $\beta_2$.
   2) Obtain $P_1$, $P_2$, and $R_{sum}$ for fixed values of $\rho_1$ and $\rho_2$ through solving problem (21) using CVX.
   3) Update r with the bisection search method: If 2) is feasible, set $r_{low} = r$; otherwise, $r_{up} = r$.
6: Until $r_{up} - r_{low} < \epsilon$, where $\epsilon$ is a small positive number. Thus we get $R_{sum}(\eta(k))$.
7: $k = k + 1$
8: Obtain $R_{sum}(\eta^o)$ by comparing all $R_{sum}(\eta(k)), k = 1, 2, \cdots$, Corresponding $P_1$, $P_2$ are the optimal transmit powers $P_1^*$, $P_2^*$.

---

### B. Power-Splitting Ratio Optimization

To obtain the optimal value for the received power splitter coefficients $\rho_1$ and $\rho_2$, problem (19) is reformulated taking into account the optimal transmit powers $P_1^*$ and $P_2^*$ as

$$\max_{\rho_1, \rho_2 \in (0,1)} R_{sum} \quad \text{s.t.} \tag{22a}$$
$$Q_{1 \to 2} \geq \bar{Q}_{1 \to 2} \tag{22b}$$
$$Q_{2 \to 1} \geq \bar{Q}_{2 \to 1}. \tag{22c}$$

Clearly, from equation (16) and (17), the received power splitter coefficients $\rho_1$ and $\rho_2$ are separable with respect to the objective functions and constraints in problem (22). Hence, we can decompose problem (22) into two sub-problems, namely,

$$\max_{\rho_1 \in (0,1)} C_{2 \to 1} \quad \text{s.t.}$$
$$Q_{2 \to 1} \geq \bar{Q}_{2 \to 1} \tag{23}$$

and

$$\max_{\rho_2 \in (0,1)} C_{1 \to 2} \quad \text{s.t.}$$
$$Q_{1 \to 2} \geq \bar{Q}_{1 \to 2} \tag{24}$$

for optimizing $\rho_1$ and $\rho_2$, respectively.

Let us first analyze the case of optimizing $\rho_1$. We obtain the Lagrangian of problem (23) as

$$\mathcal{L}(\rho_1, \lambda_1)$$
$$= \log_2\left(\frac{\rho_1(P_2^*|h_{cb}|^2 + (|\hat{h}_{ab}|^2 + \epsilon_1^2)P_1^* + \sigma_{A_1}^2) + \sigma_{p_1}^2}{\rho_1((|\hat{h}_{ab}|^2 + \epsilon_1^2)P_1^* + \sigma_{A_1}^2) + \sigma_{p_1}^2}\right)$$
$$+ \lambda_1[(1-\rho_1)(|h_{cb}|^2 P_2^* + (|\hat{h}_{ab}|^2 - \epsilon_1^2)P_1^* + \sigma_{A1}^2)$$
$$+ \bar{Q}_{2\to 1}] \qquad (25)$$

where $\lambda_1 \geq 0$ is the Lagrangian multiplier associated with the energy harvesting constraint. Obtaining the first order derivative $\frac{\partial \mathcal{L}}{\partial \rho_1}$ of (25) and after performing some mathematical manipulations, we have

$$\frac{\partial \mathcal{L}(\rho_1, \lambda_1)}{\partial \rho_1} = \frac{P_2^*|h_{cb}|^2 \sigma_{p_1}^2}{(\rho_1((|\hat{h}_{ab}|^2 + \epsilon_1^2)P_1^* + \sigma_{A_1}^2) + \sigma_{p_1}^2)^2}$$
$$- \lambda_1(|h_{cb}|^2 P_2 + (|\hat{h}_{ab}|^2 - \epsilon_1^2)P_1 + \sigma_{A_1}^2). \qquad (26)$$

The Lagrangian dual variable $\lambda_1$ is selected such that the energy harvesting constraint in (23) is satisfied to equality. After some algebraic manipulation, the following second-order polynomial is obtained from (26)

$$a_1 \rho_1^2 + \rho_1 b_1 + c_1 = 0 \qquad (27)$$

where $a_1 = ((|\hat{h}_{ab}|^2 + \epsilon_1^2)P_1^* + \sigma_{A_1}^2)^2$, $b_1 = 2a_1\sigma_{p_1}^2$, $c_1 = (\sigma_{p_1}^2)^2 - \frac{P_2^*|h_{cb}|^2 \sigma_{p_1}^2}{\lambda_1 D_1}$, and $D_1 = |h_{cb}|^2 P_2^* + (|\hat{h}_{ab}|^2 + \epsilon_1^2)P_1^* + \sigma_{A_1}^2$. Since the transmit power must be non-negative, the only acceptable solution of equation (27) is given by

$$\rho_1^* = \frac{-b_1 + \sqrt{b_1^2 - 4a_1 c_1}}{2a_1}. \qquad (28)$$

Note that $\rho_1^*$ is an increasing function of the Lagrangian multiplier $\lambda_1$ which must be chosen such that

$$\sqrt{b_1^2 - 4a_1 c_1} \geq b_1. \qquad (29)$$

Equation (29) can be further simplified as $4a_1 c_1 \leq 0$. Since $a_1 = ((|\hat{h}_{ab}|^2 + \epsilon_1^2)P_1^* + \sigma_{A_1}^2)^2 > 0$ always holds in practice, we can conclude that $c_1 \leq 0$. Thus from the definition of $c_1$, we obtain

$$(\sigma_{p_1}^2)^2 - \frac{P_2^*|h_{cb}|^2 \sigma_{p_1}^2}{\lambda_1 D_1} < 0. \qquad (30)$$

From (30), we define the upper bound of $\lambda_1$ as

$$\lambda_1 \leq \frac{P_2^*|h_{cb}|^2 \sigma_{p_1}^2}{\sigma_{P_1}^4 D}.$$

Now we can search for the optimal $\lambda_1$ within the following interval

$$\frac{P_2^*|h_{cb}|^2}{\sigma_{p_1}^2 D} \geq \lambda_1 \geq 0.$$

Similar results can be derived for optimal $\rho_2$ in problem (24) as

$$\rho_2^* = \frac{-b_2 + \sqrt{b_2^2 - 4a_2 c_2}}{2a_2}, \qquad (31)$$

where we define $a_2 = ((|\hat{h}_{cd}|^2 + \epsilon_2^2)P_2^* + \sigma_{A_2}^2)^2$, $b_2 = 2a_2 \sigma_{p_2}^2$, $c_2 = (\sigma_{p_2}^2)^2 - \frac{P_1^*|h_{ad}|^2 \sigma_{p_2}^2}{\lambda_2 D_2}$, and $D_2 = |h_{ad}|^2 P_1^* + (|\hat{h}_{cd}|^2 + \epsilon_2^2)P_2^* + \sigma_{A_2}^2$.

### C. Iterative Update

Now, the original transmit power and receive power splitter optimization problem (19) can be solved by an iterative technique shown in Algorithm 2. Algorithm 2 continually updates the objective function until convergence. Note that the constraints in problem (19) are always satisfied in every update as long as the condition on the choice of $\lambda$ is met.

---
**Algorithm 2** Procedure for solving problem (19)
---
1: Initialise $\rho_1$ and $\rho_2$.
2: Repeat
   1) Solve subproblem (21) using Algorithm 1 to obtain optimal $P_1$ and $P_2$.
   2) Solve subproblems (23) and (24) to obtain optimal $\rho_1$ and $\rho_2$.
3: Until convergence.

---

## IV. NUMERICAL EXAMPLES

In this section, we study the performance of the proposed transmit power and received power splitting optimization algorithm for SWIPT in FD communication systems through numerical simulations. We simulate a flat Rayleigh fading environment where the channel fading coefficients are characterized as complex Gaussian numbers having entries with zero mean and are independent and identically distributed. For simplicity, we assume $\bar{Q}_{1\to 2} = \bar{Q}_{2\to 1} = Q$ and $\rho_1 = \rho_2 = \rho$ unless explicitly mentioned otherwise. All simulations are averaged over 1000 independent channel realizations.

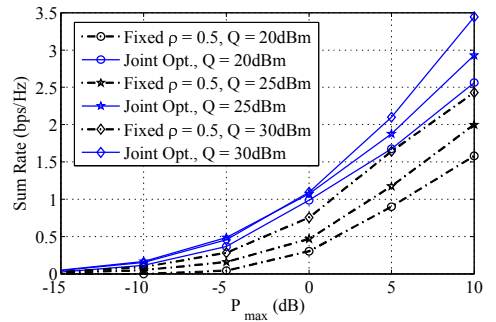

Fig. 2. Sum-rate versus P$_{\max}$.

In Fig. 2, we investigate the sum-rate performance of the proposed algorithm versus transmit power budget P$_{\max}$ (dB) for different values of the harvested energy constraint. In particular, sum-rate results of the proposed joint transmit power and receive power splitter optimization scheme ('Joint Opt.' in the figures) is compared in Fig. 2 with those of the transmit power only optimization scheme (given $\rho$). For simplicity, we assume that 70% of the self-interference power has been cancelled using existing analog and digital cancellation techniques [6]. Interestingly, the proposed joint optimization scheme yields noticeably higher sum-rate compared to the transmit power only optimization which essentially necessitates joint optimization. It can also be observed that as we increase $P_{\max}$, the sum-rate rate for both schemes increases. Also, the increased harvested energy constraints demand more power to be transmitted and hence yields higher sum-rate.

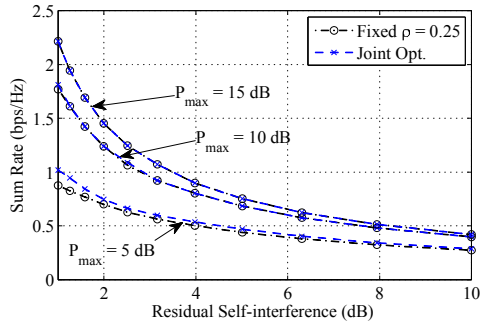

Fig. 3. Sum-rate versus residual self-interference above noise power

In the last figure, we analyze the impact of the residual self-interference on the sum-rate. Particularly, we investigate the performance of our proposed scheme for both fixed PS and joint optimization versus the residual self-interference (dB) above noise level for different values of transmit power constraint. We can see from Fig. 3 that an increase in the residual self-interference results in a corresponding decrease in the achievable sum-rate. Also, we see that the sum-rate decreases faster at higher transmit power.

## V. Conclusion

In this paper, we have investigated SWIPT in FD communication systems and proposed transmit power and received power splitting ratio optimization algorithms which maximise the sum-rate subject to the transmit power and harvested energy constraints. Using the rate-split method, we developed an iterative algorithm to optimize transmit power and receive power splitters which maximise the sum-rate of SWIPT in FD. We showed through simulation results that the residual self-interference, if not properly handled, inhibits system performance, thus reducing the achievable sum-rate.

## Acknowledgment

The author would like to thank the federal republic of Nigeria and the implementation committee of the presidential special scholarship scheme for innovation and development (PRESSID) for providing PhD research funds.